\documentclass[twocolumn,showpacs,superscriptaddress,amsmath,aps]{revtex4}
\usepackage{graphicx,color}
\usepackage{CJK}
\usepackage{bm}
\usepackage[hypertex]{hyperref}
\usepackage{float}

\newcommand{\be}{\begin{equation}}
\newcommand{\ee}{\end{equation}}
\newcommand{\bea}{\begin{eqnarray}}
\newcommand{\eea}{\end{eqnarray}}
\newcommand{\bsube}{\begin{subequations}}
\newcommand{\esube}{\end{subequations}}

\newcommand{\Eq}[1]{Eq.\,(\ref{#1})}

\newcommand{\ra}{\rangle}

\newcommand{\nl}{\nonumber \\}

\newcommand{\beq}{\begin{equation}}
\newcommand{\eeq}{\end{equation}}
\newcommand{\beqn}{\begin{eqnarray}}
\newcommand{\eeqn}{\end{eqnarray}}
\newcommand{\bsub}{\begin{subequations}}
\newcommand{\esub}{\end{subequations}}

\begin{document}
\begin{CJK*}{GBK}{Song}

\title{Estimation of parameters in circuit QED
       by continuous quantum measurement}

\author{Cheng Zhang}
\affiliation{Department of Physics, Beijing Normal University, Beijing 100875, China}

\author{Kai Zhou}
\affiliation{Department of Physics, Beijing Normal University, Beijing 100875, China}

\author{Wei Feng}
\affiliation{Center for Joint Quantum Studies,
School of Science, Tianjin University, Tianjin 300072, China}

\author{Xin-Qi Li}
\email{xinqi.li@tju.edu.cn}
\affiliation{Center for Joint Quantum Studies,
School of Science, Tianjin University, Tianjin 300072, China}
\affiliation{Department of Physics, Beijing Normal University, Beijing 100875, China}

\begin{abstract}
Designing high-precision and efficient protocols
is of crucial importance
for quantum parameter estimation in practice.
Estimation based on continuous quantum measurement is
one possible type of this, which also appears to be the most natural choice
{ for} continuous dynamical process{es}.
In this work we consider the state-of-the-art
superconducting circuit quantum-electrodynamics (QED) systems,
where high-quality continuous measurements have been
extensively performed in the past decade.
Within the framework of Bayesian estimation and particularly
using the quantum Bayesian rule in circuit QED,
we numerically simulate the likelihood function as estimator
for the Rabi frequency of qubit oscillations.
We find that, by proper design of the interaction strength of measurement,
the estimate precision can scale with the measurement time
{\it beyond} the standard quantum limit, which is usually assumed
for this type of continuous measurement.
This unexpected result is
supported by the simulated Fisher information,
and can be understood as a consequence of the quantum correlation
between the output signals by simulating the effect of quantum
efficiency of measurement.
\end{abstract}

\maketitle

\section{Introduction}

The problem of accurately estimating unknown parameters
is of both
theoretical interest and of practical importance \cite{He69,Ho11}.
In order to minimize the estimation uncertainties,
a variety of strategies have been developed in the science
of quantum metrology over the past decades \cite{GLM04,GLM11,Dem15}.
In this context, a central topic is how to exploit quantum techniques
to achieve parameter estimation with precision beyond
that obtainable by any classical scheme \cite{GLM04,GLM11,Dem15}.
For example, if a system is initially prepared in a spin
coherent state,
the precision of frequency estimation scales with
the total spin number $N$, as $1/\sqrt{N}$
which is referred to standard-quantum-limit (SQL).
However, if extra quantum resources such as
spin squeezing or entanglement are exploited,
an enhanced precision can be achieved approaching
the ultimate Heisenberg-limit (HL) scaling ($\sim1/N$) \cite{Win92,Win96}.

For the theory of quantum parameter estimation,
the concept of quantum Fisher information \cite{Hels76,Cav94}
and the associated quantum Cram\'er-Rao bound (CRB) \cite{Cra46,Rao45}
have been developed to set the minimum variance
for unbiased estimation strategies based on measurements.
However, it is not obvious how to design appropriate/optimal scheme of measurement.
For different measurement schemes, which is usually characterized
by a specific POVM
or estimator, the associated Fisher Information would set different bound
of precision according to the Cram\'er-Rao inequality.
Designing high-precision and efficient scheme of measurement is
thus of crucial importance for parameter estimation in practice.

Owing to the practical use and the rich underlying physics,
in the past years there have been considerable interests in the
quantum estimation of parameters using the output signals
of continuous measurement
\cite{Jac11,Molm13,Molm14,Molm15,Molm16,Jor17,Par17,Gen17}.
This is also the most natural choice for
parameter estimation in continuous dynamical process.
This scheme has the obvious advantage of a high efficiency ---
unlike the conventional {\it ensemble} measurement in quantum theory,
it does not need to generate the identical copies of quantum system
in order to extract meaningful results from measurements.
Following the seminal work \cite{Jac11}, the subsequent studies by M\o lmer {\it et al}
\cite{Molm13,Molm14,Molm15,Molm16} formulated the parameter estimation
based on continuous measurement as a Bayesian scheme,
for the specific example of fluorescence radiation from two-level atoms.
More recently, the same problem was investigated with a focus on
how to speed up the estimation by avoiding numerically
integrating the stochastic master equation \cite{Jor17}.
This type of parameter estimation has also been considered to include
the technique of quantum smoothing \cite{MT09a,MT09b,MT10,MT15},
as a generalization of the classical signal smoothing.

In the present work, we extend the research further to the superconducting
circuit quantum-electrodynamics (QED) system \cite{Bla04,Sch04,Sch08},
which is one of the leading platforms for quantum information processing
and for quantum measurement and control studies.
Particularly, sound studies have been performed for continuously
tracking the stochastic evolution of the qubit state in this system,
say, tracking the so-called quantum trajectories (QT)
\cite{DiCa13,Dev13a,Dev13,Sid13,Sid15,Sid12}.
On the theoretical side,
the quantum Bayesian rule has been well developed for circuit QED
in the past years \cite{Kor11,Kor16,Li14,Li16,Li18}, in some cases
promising the advantage of being more efficient
than numerically integrating the quantum trajectory equation \cite{Gam08,WM09,Jac14}.
Therefore, within the framework of Bayesian parameter estimation
\cite{Molm13,Molm14,Molm15,Molm16,Jor17},
it seems a perfect choice for us to employ the quantum Bayesian rule of circuit QED
for state update associated with the continuous measurement.
For parameter estimation, the quantum Bayesian approach can make
the calculation of the associated likelihood function
very straightforward and quite efficient \cite{Jor17}, i.e.,
using the accumulated output currents over relatively large time interval.

In this work we perform direct simulation for large number
of estimations to extract the statistical errors.
Our simulation is based on the realistic (and `standard')
dispersive readout in circuit QED system.
We investigate the effects of measurement strength, measurement time
(processing time $T$ of data collection),
and quantum efficiency of the measurement. As expected,
we find that the estimate precision is improved with increasing $T$.
However, interesting and surprisingly, we find that
by proper adjustment of the measurement strength,
the estimate precision can exceed the standard quantum limit,
manifesting a scaling behavior with $T$ in between
the SQL ($1/\sqrt{T}$) and HL ($1/T$).
This result differs from what has been assumed for similar estimation
of this type \cite{Molm13,Molm14,Molm15,Molm16,Jor17},
where the SQL scaling was concluded.
We attribute the reason for our result to {\it quantum correlation}
between the output signals of the measurement \cite{Leg85,Pala10}.
This type of correlation in time shares some nature
with the quantum entanglement \cite{Pala10},
while the latter (as a unique quantum resource)
can usually result in precision better than SQL.
We may also relate this understanding with the hints from extreme cases
such as vanishing probe interaction \cite{Molm14,Molm16,Gut11}
and dynamical phase transition \cite{Gar13,Gar16},
which can result in precision with Heisenberg scaling
owing to the quantum correlation in time.

\section{Bayesian Rule in circuit QED}

Let us consider a superconducting qubit
coupled to a waveguide cavity, i.e., the circuit-QED architecture.
In the dispersive regime, the qubit-cavity interaction is
well described by the Hamiltonian \cite{Bla04,Sch04},
$H_{\rm int}= \chi a^{\dagger}a\sigma_z$,
where $\chi$ is the dispersive coupling strength,
$a^{\dagger}$ and $a$ are the
creation and annihilation operators of the cavity mode,
and $\sigma_z$ is the qubit Pauli operator.
Associated with single quadrature homodyne measurement
for microwave transmission/reflection,
the output current can be reexpressed as (after the so-called polaron
transformation to eliminate the degrees of freedom of the cavity photons)
\cite{Li14,Li16,Li18,Gam08}
\bea\label{I-t}
I(t)=-\sqrt{\Gamma_{ci}(t)}\langle \sigma_{z}\rangle+\xi(t).
\eea
In this result, $\xi(t)$,
originated from the fundamental quantum-jumps,
is a Gaussian white noise and satisfies the ensemble-average property
$E[\xi(t)]=0$ and $E[\xi(t)\xi(t')]=\delta(t-t')$.
$\Gamma_{ci}(t)$ is the coherent information gain rate which,
together with, say, the no-information back-action rate $\Gamma_{ba}(t)$
and the overall measurement decoherence rate $\Gamma_d(t)$, is given by
\begin{subequations}\label{GM3t}
\begin{eqnarray}
  && \Gamma_{ci}(t) = \kappa|\beta(t)|^2\cos^2(\varphi-\theta_\beta) \,,
   \\
  &&  \Gamma_{ba}(t) = \kappa|\beta(t)|^2\sin^2(\varphi-\theta_\beta)\,,
  \\
  &&  \Gamma_d(t)=4\chi\mathrm{Im}[\alpha^*_1(t)\alpha_2(t)] \,.
\end{eqnarray}
\end{subequations}
Here $\varphi$ is the local oscillator's (LO)
phase in the homodyne measurement,
$\kappa$ is the leaky rate of the microwave photon from the cavity,
and $\beta(t)=\alpha_2(t)-\alpha_1(t)
\equiv |\beta(t)|e^{i\theta_{\beta}}$
with $\alpha_1(t)$ and $\alpha_2(t)$ the cavity fields
associated with the qubit states $|1\ra$ and $|2\ra$, respectively.

More detailed discussions of the physical meanings of the above rates
can be found in Refs.\ \cite{Li14,Li16,Li18,Gam08}.
Briefly speaking, the information-gain rate $\Gamma_{ci}$
is associated with inferring the qubit state $|e\ra$ or $|g\ra$
from the output current of measurement,
while $\Gamma_{ba}$ characterizes the backaction of measurement
not associated with the qubit-state-information gain,
but rather with the qubit-level fluctuations.
$\Gamma_{d}$ corresponds to the overall decoherence rate after
ensemble-averaging a large number of quantum trajectories.
The sum of the former two rates,
$\Gamma_m=\Gamma_{ci}+\Gamma_{ba}$, is the total measurement rate.
If $\Gamma_m=\Gamma_d$, the measurement is ideally quantum limited,
with quantum efficiency $\eta=\Gamma_m/\Gamma_d=1$.
Otherwise, if $\Gamma_m<\Gamma_d$, the measurement is not {\it ideal},
implying some information loss.

In steady state, the cavity fields read
\bea\label{ss-field}
\bar{\alpha}_{1,2}=-\epsilon_m [(\Delta_r \pm \chi)-i \kappa/2]^{-1} \,,
\eea
where $\Delta_r$ is the frequency offset between the measuring microwave
(with amplitude $\epsilon_m$) and the cavity mode.
In this work, rather than considering a general set-up of the circuit-QED system
\cite{Kor16,Li14,Li16}, we restrict considerations
to the bad-cavity and weak-response limits.
Under this condition, the transient process
of $\alpha_1(t)$ and $\alpha_2(t)$ is not important.
All the rates shown above can be calculated with the
steady-state fields $\bar{\alpha}_{1,2}$ given by \Eq{ss-field}.

Corresponding to the qubit state $|1\ra$ ($|2\ra$) and
after averaging the continuous current over time interval $\tau$, i.e.,
${\cal I}_m=(1/\tau) \int^{t+\tau}_{t} dt' I(t')$,
the coarse-grained output current ${\cal I}_m$
is a stochastic variable centered at
$\bar{I}_{1(2)}=\mp \sqrt{\Gamma_{ci}}$
and satisfies the Gaussian distribution with probability
\bea\label{G-P12}
P_{1(2)}(\tau)  = (2\pi V)^{-1/2}
\exp\left[-({\cal I}_m-\bar{I}_{1(2)})^2  / (2V) \right],
\eea
where $V=1/\tau$ is the distribution variance.

Now consider an arbitrary quantum superposed state $\rho(t)$ (at the moment $t$).
Based on the subsequent (coarse-grained) current $I_m$,
the quantum Bayesian rule updates the qubit state as follows
\cite{Kor11,Kor16,Li14,Li16}. For the diagonal elements,
\begin{subequations}\label{BR-cQED}
\bea\label{BR-cQED-b}
\rho_{jj}(t+\tau) = \rho_{jj}(t) \, P_j (\tau)/ {\cal N}(\tau) \,,
\eea
where $j=1,2$ and ${\cal N}(\tau)=\rho_{11}(t)P_1(\tau)+\rho_{22}(t)P_2(\tau)$.
This is nothing but the Bayes' theorem in probability theory.
For the off-diagonal elements, which are unique in quantum theory,
\bea\label{BR-cQED-a}
&& \rho_{12}(t+\tau) = \rho_{12}(t)
   \left[\sqrt{P_1(\tau)P_2(\tau)}/{\cal N}(\tau) \right]  \nl
&& ~~~~~~ \times D(\tau) \,
\exp\left\{-i[\Phi_1(\tau)+ \Phi_2(\tau)]\right\}  \,.
\eea
\end{subequations}
In this result, the purity factor reads
$D(\tau)= e^{-(\Gamma_d-\Gamma_m)\,\tau/2}$.
We remind the reader that the measurement rate is given by
$\Gamma_m=\Gamma_{ci}+\Gamma_{ba}$.
Using the steady-state solutions, \Eq{ss-field},
one can easily prove that $\Gamma_d=\Gamma_m$.
Thus, in the bad-cavity limit (no transient dynamics of the cavity field),
the {\it intrinsic} $D$-factor in the successive Bayesian update
can be approximated by unity.
In order to account for decoherence of {\it external} origins
(such as photon loss and/or amplifier's noise),
one can simply introduce an extra rate $\Gamma_{\varphi}$,
thus $D(\tau)= e^{-\Gamma_{\varphi}\,\tau/2}$.

The first phase factor in \Eq{BR-cQED-a}, $e^{-i\Phi_1(\tau)}$,
is associated with an ac-Stark-shift modified unitary phase accumulation,
i.e., with $\Phi_1(\tau)=(\Omega_q+B)\tau$
where the ac-Stark-shift of the qubit energy ($\Omega_q$)
reads $B=2\chi{\rm Re}(\bar{\alpha}_1 \bar{\alpha}^*_2)$.
Of more interest is the second phase factor
$e^{-i\Phi_2(\tau)}=e^{i\Gamma_{ba}(I_m\tau)}$,
which is associated with the accumulated random `charge'
and reflects the no-information gain backaction on the qubit.
For a more detailed discussion on this stochastic phase factor
the reader is referred to Refs.\ \cite{Kor11,Kor16,Li14,Li16}.

\section{Method}

We assume now that the superconducting qubit is subject to a Rabi drive
and at the same time subject to continuous measurement.
Our goal is to estimate the Rabi frequency
from the output current of the continuous measurement.
The stochastic evolution of the qubit (the quantum trajectory)
is governed by the following iterative rule
\bea\label{UM-j}
\rho(t_j) = {\cal U}_j{\cal M}_j[\rho(t_{j-1})] \,,
\hspace{0.2cm} {\rm with}~~ j=1,2, \cdots N \,.
\eea
Here we have discretized the evolution with time interval $\tau$,
with thus a total measurement time $T=N\,\tau$.
The superoperator ${\cal M}_j$ accounts for the measurement-induced change
of the qubit state, whose performance is explicitly given
by the quantum Bayesian rule.
The superoperator ${\cal U}_j$ in \Eq{UM-j}
describes the unitary evolution caused by the Rabi drive, i.e.,
${\cal U}_j(\cdots)=e^{-i{\cal L}_q\tau}(\cdots)
= e^{-i\tilde{H}_q\tau}(\cdots)e^{i\tilde{H}_q\tau}$
with $\tilde{H}_q$ the qubit Hamiltonian under Rabi drive
and renormalized by the measurement (i.e. with the ac-Stark shift).
For small $\tau$, the action order of ${\cal U}_j$
and ${\cal M}_j$ is irrelevant.

Based on the rule of \Eq{UM-j}, we know the qubit state $\rho(t_j)$
after the $j_{\rm th}$ step evolution, conditioned on
the coarse-grained current ${\cal I}_j$.
Meanwhile, for this $j_{\rm th}$ step of measurement,
the probability of getting ${\cal I}_j$ is
\bea\label{P-1}
{\cal P}({\cal I}_j)=\rho_{11}(t_{j-1})P_1(\tau)+ \rho_{22}(t_{j-1})P_2(\tau) \,,
\eea
with $P_1(\tau)$ and $P_2(\tau)$ given by \Eq{G-P12}.
Then, straightforwardly, the {\it joint probability} of getting
the results $\{{\cal I}_1,{\cal I}_2, \cdots {\cal I}_N \}$
is simply a product of the individual probabilities
\bea\label{P-2}
{\cal P}(\{{\cal I}_1,{\cal I}_2, \cdots {\cal I}_N\}|\Omega)
=  \prod^{N}_{j=1} {\cal P}({\cal I}_j)  \,.
\eea
Here we explicitly indicate that this probability depends on
the parameter $\Omega$ (the {\it possible} Rabi frequency).

We expect, from simple intuition, that the true Rabi frequency $\Omega_R$
will be {\it most compatible} with the output results
$\{{\cal I}_1,{\cal I}_2, \cdots {\cal I}_N \}$,
leading thus to maximum probability.
Therefore, it is plausible that we get an estimate value $\Omega_{ML}$
for $\Omega_R$ from the location of the maximum
of the probability function
${\cal P}(\{{\cal I}_1,{\cal I}_2, \cdots {\cal I}_N\}|\Omega)$,
referred to in the literature as the {\it likelihood function}.
Using a different $\Omega$ (rather than $\Omega_R$) to calculate
${\cal P}(\{{\cal I}_1,{\cal I}_2, \cdots {\cal I}_N\}|\Omega)$,
based on Eqs.\ (\ref{UM-j})-(\ref{P-2}),
should result in a smaller probability.
This constitutes the basic idea of the
{\it maximum-likelihood-estimation} (MLE) method.

Essentially, the MLE method is a Bayesian approach for parameter estimation.
One may imagine, to start with,
a uniform distribution ${\cal P}(\Omega)$
over a certain range. The uniform distribution means that
we have no knowledge about $\Omega_R$.
After getting the data record of measurement and performing the Bayesian inference,
the knowledge about $\Omega_R$ changes to a new probability
${\cal P}(\Omega|{\cal I}_1,\cdots,{\cal I}_N)$.
The peak of this new distribution can also be an estimate for $\Omega_R$,
which should correspond to the estimated value $\Omega_{ML}$ from the MLE method.

In practice, the following {\it log-likelihood function} is
used for the parameter estimation
\bea
L(\Omega)=\ln\, {\cal P}(\{{\cal I}_1,{\cal I}_2, \cdots {\cal I}_N\}|\Omega) \,,
\eea
in order to make the maximum peak more prominent.
In Fig.\ 1, we plot this function to illustrate the MLE method
(using {\it dimensionless units} here and in remainder of this work).
$L(\Omega)$ is computed using the single realization of
continuous measurement current $I(t)$ over $(0,T)$,
by coarse-graining it into
$\{{\cal I}_1, {\cal I}_2,\cdots,{\cal I}_N  \}$ with $N=T/\tau$.
Notice that this splitting can be rather arbitrary, i.e.,
with $L(\Omega)$ not influenced by the choice of $\tau$.
The only requirement is that $\tau$ should not be too large to violate
the precision of the Bayesian update (in the presence of Rabi oscillation).
In the whole simulations of this work, we choose $\tau=1000\, dt=10^{-3}\tau_R$,
while the time increment $dt=10^{-6}\tau_R$ is set
for simulating the quantum trajectory equation \cite{Li14,Li16,Li18,Gam08}
to generate the continuous output current.
Through the whole work, the Rabi period ($\tau_R=2\pi/\Omega_R$)
is used as the units of time.
Again, we mention that the $\Omega$ dependence of $L(\Omega)$
is introduced through the unitary operator
$e^{-i{\cal L}_q\tau}$ in each step of state update.

\begin{figure}[!htbp]
  \centering
  \includegraphics[width=7.5cm]{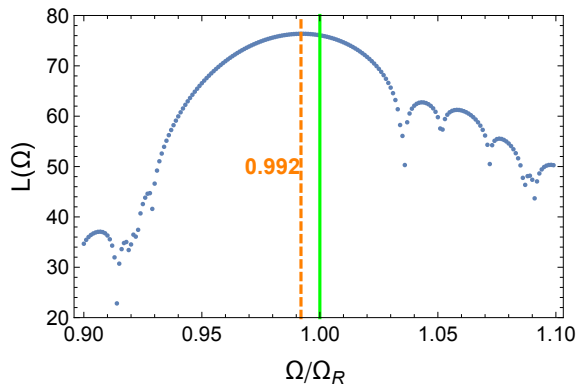}
  \caption{
Illustration of the estimation method. The likelihood function
is calculated (with un-normalized probability functions
$P_{1,2}(\tau)$ in Eq.\ (7), `separated' from Eq.\ (4))
using the quantum Bayesian rule
by choosing the coarse-grained time $\tau=1000\, dt=10^{-3}\tau_R$,
while $dt=10^{-6}\tau_R$
is set for simulating the quantum trajectory equation
to generate the continuous output current (for time $T=N\,\tau$ with $N=10^5$).
Through the whole work we assume the {\it true} Rabi frequency $\Omega_R/2\pi=1.0$
(in arbitrary dimensionless units),
and accordingly we use the Rabi period $\tau_R=2\pi/\Omega_R$ as the units of time.
In this plot we set the measurement strength $\Gamma_m=0.25\Omega_R$
and obtain the estimated value $\Omega_{ML}=0.992\Omega_R$
from the position of the maximum of the peak.}    
\label{LF}
\end{figure}

We consider a resonant Rabi drive with {\it true} Rabi frequency
$\Omega_R/2\pi=1$ (in arbitrary dimensionless units).
In the present proof-of-principle simulation, we assume $\Delta_r=0$
(thus $\theta_{\beta}=0$) and consider the maximal information gain
with LO phase $\varphi=0$.
Therefore we have $\Gamma_{ba}=0$, $\Gamma_{m}=\Gamma_{ci}$
and $\Gamma_{d}=\Gamma_m$ (owing to the bad-cavity limit).
Except for the data shown in Fig.\ 4, we also do not
account for any external decoherence
in our simulation (setting $\Gamma_{\varphi}=0$).

Indeed, as shown in Fig.\ 1, we get an estimation for the Rabi frequency
at $\Omega_{ML}=0.992\Omega_R$, from the maximum peak position of $L(\Omega)$.
In this plot, we only show the log-likelihood function for a
relatively small range of $\Omega$, indicating that
we already have some prior knowledge about $\Omega_R$.
If we had poor knowledge about $\Omega_R$,
we should calculate $L(\Omega)$ for a wider range.
In this case, more peaks may appear in $L(\Omega)$.
An even worse situation would arise if the maximum peak
would not occur near $\Omega_R$.
This would imply a failure of the estimation and the result should
be discarded.

Another point is that, in order to get a convergent estimation,
one should sample a relatively large number of currents
$\{{\cal I}_1, {\cal I}_2,\cdots,{\cal I}_N  \}$,
i.e., with large $N$ or more precisely large $T$
by noting that $T=N\tau$.
Actually, it has been noted that the MLE result can saturate
the Cram\'er-Rao bound when $N$ is large enough
\cite{Molm13,Molm14,Molm15,Molm16}.
However, the classical Cram\'er-Rao bound is determined by
the classical Fisher information which is associated with
specific schemes of measurement.
It has been well understood that the more sensitive dependence of
the output results on the parameter will result in better precision.
Searching for an optimal measurement protocol in practice
is thus of crucial importance but is unclear in general.
In the following, in Fig.\ 2, we will further discuss this point.

A final remark is that possible quantum correlation effect
may be contained in the likelihood function $L(\Omega)$.
This is in some sense similar to the reason of violating the Leggett-Garg
inequality (a type of Bell's inequality in time) \cite{Leg85} as demonstrated
in this same circuit-QED system via continuous measurements \cite{Pala10}.
We will come back to this point later after displaying the result
beyond the standard quantum limit.

\section{Results}

To characterize the estimation errors,
we introduce the root-mean-square (RMS) variance
\bea
\delta\Omega= \left(
\sum^{M}_{k=1}(\Omega^{(k)}_{ML}-\bar{\Omega}_{ML})^2\,/\,M
\right)^{1/2} \,,
\eea
where $\bar{\Omega}_{ML} = \frac{1}{M}\sum^{M}_{k=1}\Omega^{(k)}_{ML}$,
with $\Omega^{(k)}_{ML}$ the estimated result of the $k_{\rm th}$
realization based on $\{{\cal I}_1, \cdots, {\cal I}_N\}^{(k)}$.
To extract the RMS variance, we simulate $M=2000$ trajectories
for each given measurement time ($T=N\tau$).

Let us analyze the problem of {\it appropriate} measurement,
in a sense to make the measurement results
{\it more sensitive} to the parameter under estimation.
First, as mentioned above,
we should eliminate the ``realistic" (no information gain) backaction
in order to maximize the information-gain rate ($\Gamma_{ci}\to \Gamma_m$)
by adjusting the LO phase $\varphi=\theta_{\beta}=0$.
Second, we search for an {\it optimal strength} for the continuous measurement,
which can be characterized by the measurement rate $\Gamma_m$.

\begin{figure}[!htbp]
  \centering
  \includegraphics[width=6.0cm]{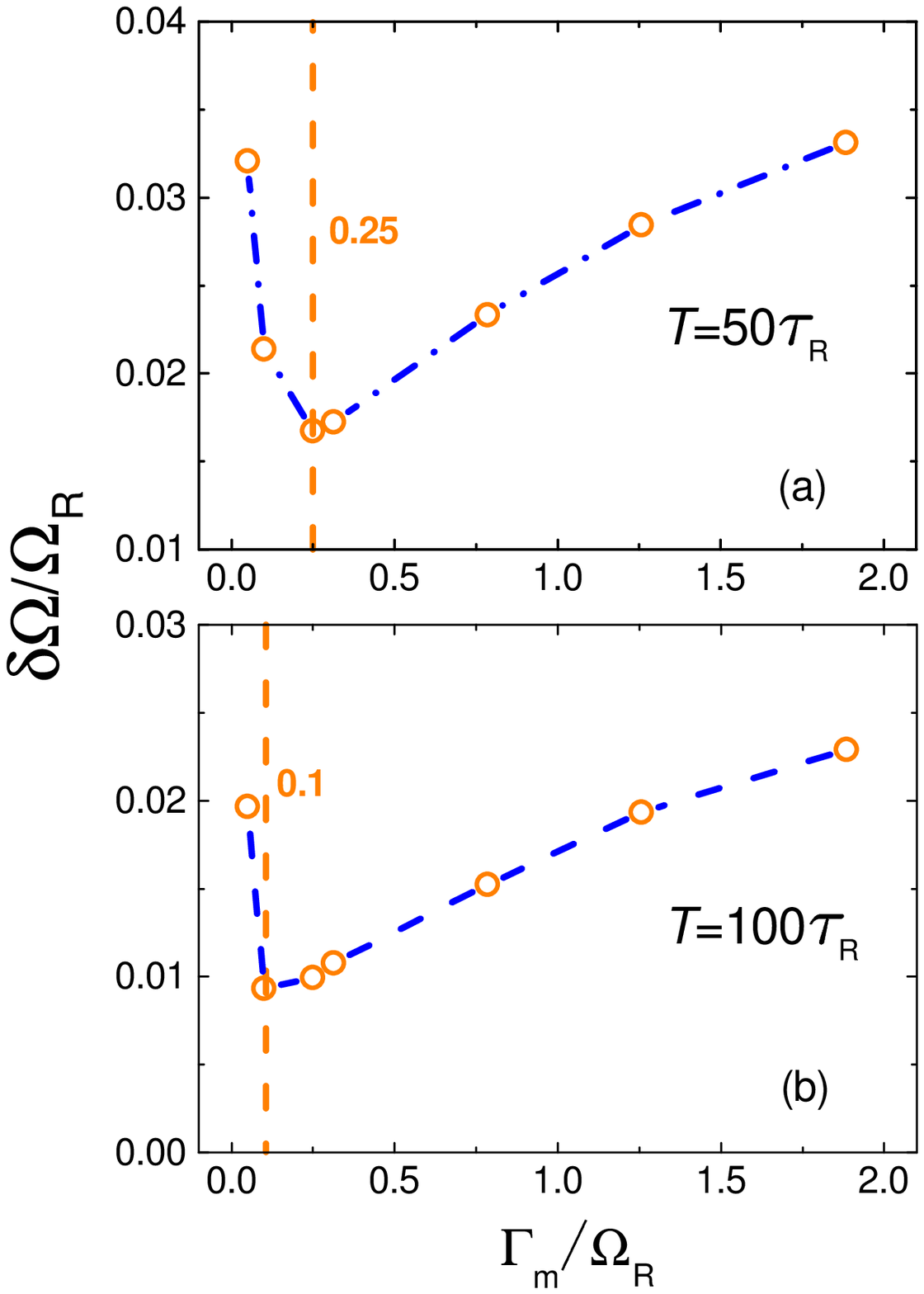}
  \caption{
Measurement strength dependence of the estimation errors (root-mean-square variance).
Owing to competition between information gain and measurement backaction,
there exists an optimal measurement strength.
The optimal strength is of $T$ dependence, as shown in (a) and (b)
for $T/\tau_R=50$ and 100 ($\tau_R$ is the Rabi period).
However, for longer $T$ as shown in (b),
the sub-optimal $\Gamma_m$ (near the optimal strength) can result in
good estimation with precision not very sensitive to $\Gamma_m$
(in the sub-optimal range).   }
  \label{optimal-strength}
\end{figure}

In Fig.\ 2(a) we show the estimation RMS variance
versus the measurement strength.
Importantly, we observe the existence of an {\it optimal} strength
of the continuous measurement. We understand the reason as follows.
From the continuous output current \Eq{I-t},
we know that for weak strength of measurement,
the noise component (the second term) will be much larger than
the information-carrying term (the first one).
In other words, the output current carries little information
of the qubit state which is governed by the parameter of Rabi frequency.
In the other extreme, for strong strength of measurement,
while the state-information-carrying component
(the first term in \Eq{I-t}) is enhanced,
the Rabi oscillation of the qubit state will be more seriously destroyed
by the measurement backaction, making thus the first term of \Eq{I-t}
not well correlated with the Rabi frequency.
That is, the enhanced strength of measurement will gradually
force the evolution into the so-called Zeno regime,
resulting in output current of telegraphic type
which is poorly correlated with the unitary Rabi drive.
Therefore, it is the competition
between the information-gain and measurement backaction
that results in the optimal measurement strength as revealed in Fig.\ 2(a).
From Fig.\ 2(b), we also find this `optimal' strength not universal,
but weakly depending on the measurement time $T$
(the size of the collected current).
For longer measurement time,
the optimal strength of measurement is smaller.
However, from Fig.\ 2(b), we see that the `sub-optimal' strength
(e.g. $\Gamma_m/\Omega_R=0.25$ rather than 0.1)
has little importance for the precision of the estimation.

\begin{figure}   
  \centering
  \includegraphics[width=5.5cm]{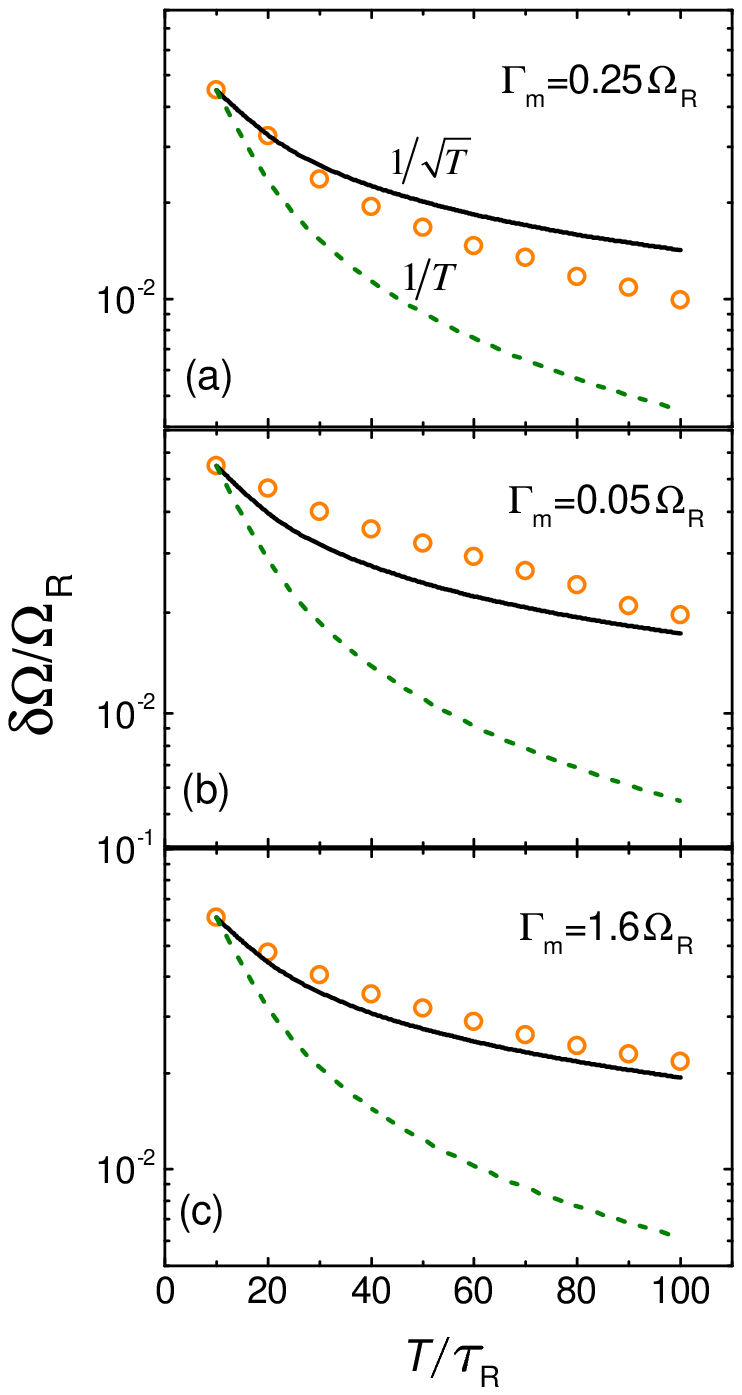}
  \caption{Estimate precision (root-mean-square variance) {\it versus} the measurement time (scaled by the Rabi period $\tau_R$).
In particular, we compare the simulated results (open circles) with the SQL
($\sim 1/\sqrt{T}$, solid line) and HL ($\sim 1/T$, dashed line) scaling behaviors.
In (a), by properly choosing the measurement strength (near the {\it optimal} one),
we find that the precision can evidently exceed the SQL.
In (b) and (c), we show that for measurement strengths deviating from the optimal/sub-optimal value, 
either the smaller or the larger values of $\Gamma_m$ cannot violate the SQL precision.} 
  \label{scaling-kappa}
\end{figure}

In quantum estimation, one of the most important problems is how
the precision scales with the `size' of the quantum resource
(e.g. the entangled photon numbers in the optical phase estimation).
For the quantum estimation based on continuous measurement
(without introducing special procedures such as feedback),
the existing studies assumed a SQL scaling
($\sim 1/\sqrt{T}$) with the measurement time $T$
\cite{Molm13,Molm14,Molm15,Molm16,Jor17}.
Below we reexamine this issue in the context of continuous
measurement in circuit-QED.
We find that, remarkably, it is possible to violate the SQL.

Let us formally denote the RMS variance as $\delta\Omega = \frac{1}{\sqrt{M}}f(T)$,
where the specific $M$ dependence is simply from the central-limit-theorem.
Our interest is to examine the $T$ dependence, especially,
to compare it with the SQL and HL scalings.
As a clear comparison, in Fig.\ 3
we compare the simulated RMS variance with the SQL
$C_1/\sqrt{T}$ (solid line)
and HL $C_2/T$ (dashed line).
Here we set the constants $C_1$ and $C_2$
by making the SQL and HL curves
coincide with the simulated RMS variance at $T=10$.
The two curves simply imply that, if the scaling is governed by SQL (HL),
the simulated results should follow the solid (dashed)
curve with increasing $T$.

In Fig.\ 3, we show results for different measurement strengths.
Remarkably, as seen in Fig.\ 3(a),
we find that by properly choosing the measurement strength
(near the {\it optimal} one), the precision can evidently exceed the SQL.
We notice that in the studies by M\o lmer {\it et al}
\cite{Molm13,Molm14,Molm15,Molm16},
only the $1/\sqrt{T}$ scaling is obtained for the Fisher information
associated with the homodyne detection for the fluorescence radiation.
This result was qualitatively understood by the measurement backaction
which results in vanished correlation between the output signals.
In the work by Jordan {\it et al} \cite{Jor17}, the $1/\sqrt{T}$ scaling
is also briefly mentioned, despite that the $T$ scaling plotted there
in Fig.\ 2(c) is a bit worse than SQL.
In Appendix A, we further support the scaling behavior in Fig.\ 3(a)
by numerically computing the Fisher information.

We may understand the result in Fig.\ 3(a)
from different perspectives as follows.
First, the `inconsistency' with Refs.\ \cite{Molm13,Molm14,Molm15,Molm16}
may originate from the different schemes of measurement.
There, the measurement operator
$\sigma_{\varphi}=\cos\varphi \sigma_x - \sin\varphi \sigma_y$
has {\it randomly flipping} backaction on the qubit.
Compared to $\sigma_z$ measurement, this type of measurement
has stronger destructive influence on the qubit,
i.e., making the population (superposition)
less associated with the Rabi frequency.

Second, for the continuous $\sigma_z$ measurement of the Rabi oscillation,
quantum correlation exists between the measurement outcomes.
Actually, this type of quantum correlation has inspired the study of the
{\it Bell-inequality-in-time}, say, the Leggett-Garg inequality \cite{Leg85}.
In particular, this quantum correlation has been experimentally demonstrated
in the circuit-QED system based on
the continuous $\sigma_z$ measurement \cite{Pala10}.
Therefore, it seems that the argument of {\it vanished correlation}
in Refs.\ \cite{Molm13,Molm14,Molm15,Molm16},
leading to the $1/\sqrt{T}$ scaling,
may not apply to our situation.

Third, for the simple estimation scheme based on continuous measurement
(not involving any special techniques),
the possibility of reaching the Heisenberg limit is not ruled out
{\it (i)}
For instance, at the end of Ref.\ \cite{Molm14},
it was pointed out that the Fisher information can
scale with $T^2$ for {\it undamped} system evolution,
such as the case if the system superposition state
does not couple to environment and the measurement is
performed on the system rather than on the emitted radiation.
{\it (ii)}
In Ref.\ \cite{Gut11},
via analyzing the quantum Markov chain defined
by a sequence of successive passage of atoms through a cavity,
it was found that
the quantum Fisher information scales quadratically rather than linearly
with the number of atoms, at the limit of weak unitary interaction.
{\it (iii)}
Another example of interest is making the system
(e.g., a driven atom under photon emissions)
approach a dynamical phase transition \cite{Gar13,Gar16}.
In that case, the quantum Fisher information may become quadratic in times
shorter than the correlation time of the dynamics.
This becomes valid for all times at the point of dynamical phase transition.

\begin{figure}[!htbp]
  \centering
  \includegraphics[width=6.0cm]{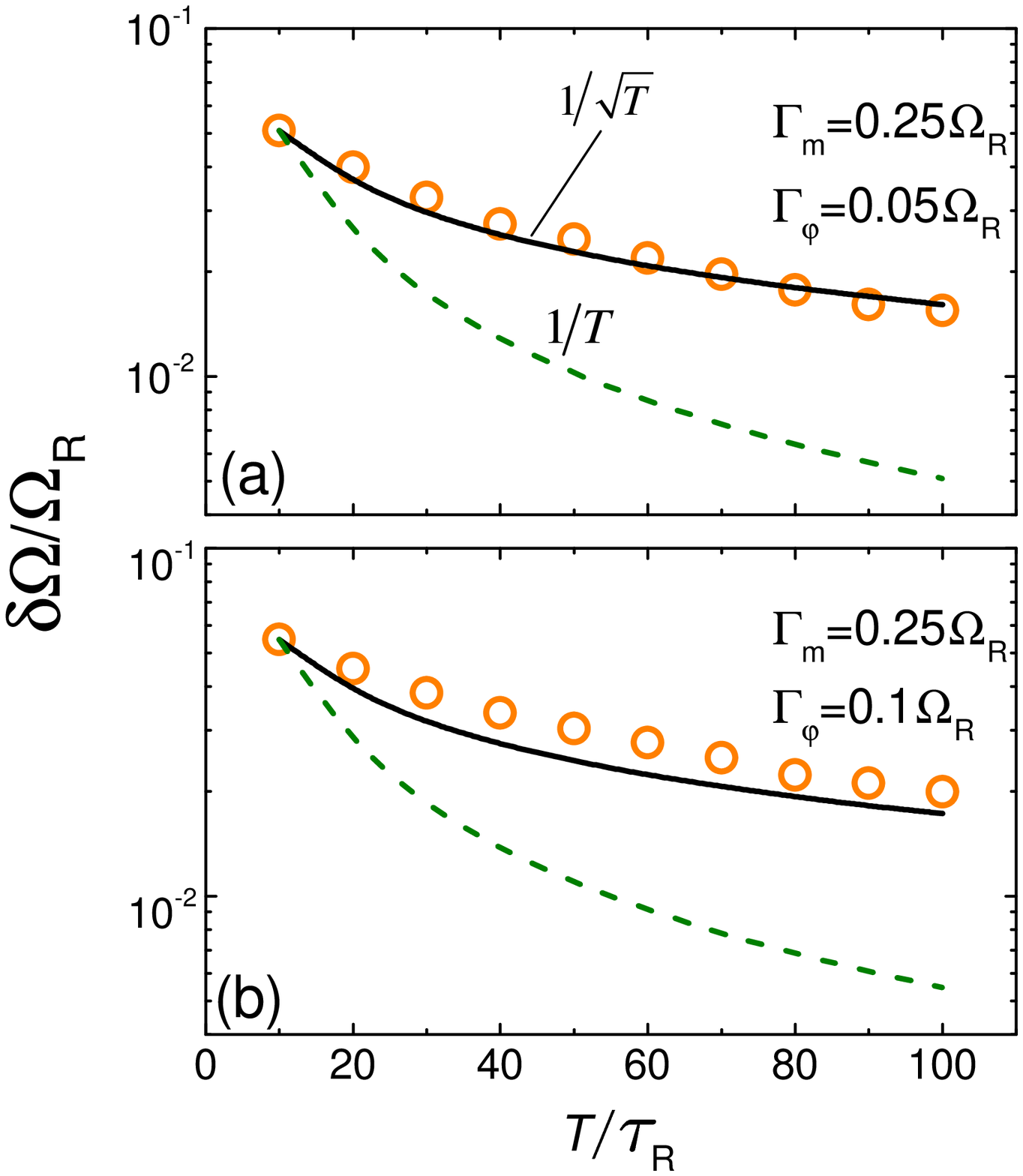}
  \caption{
Further examination of the result in Fig.\ 3(a).
Setting still the sub-optimal measurement strength $\Gamma_m=0.25\Omega_R$
but introducing extra decoherence ($\Gamma_{\varphi}$)
owing to photon loss and/or amplifier's noise during the measurement,
we find that the result can no longer exceed the SQL precision.
This supports further the understanding based on quantum correlation
to the remarkable result in Fig.\ 3(a) }
  \label{scaling-gamma}
\end{figure}

Therefore, our result in Fig.\ 3(a) does not contradict any basic physics,
but rather can fall into the category of quantum correlation.
As a trade-off between information gain and measurement backaction,
a proper strength of the continuous measurement is required:
As seen in Fig.\ 3(b) and (c),
values deviating from the optimal/sub-optimal measurement strength,
whether for smaller or larger values of $\Gamma_m$,
will not lead to a violation of
the SQL precision.
In addition to the proper measurement strength,
sufficient {\it quantum coherence}
is another condition for the result in Fig.\ 3(a).
In Fig.\ 4, we further account for the effect of {\it decoherence}
owing to non-ideal quantum measurement,
e.g., photon loss and/or amplifier's noise during the measurement.
From Fig.\ 4(a) and (b), we observe that the estimate precision
becomes worse with the increase of decoherence,
and can no longer violate the scaling of SQL
by varying the measurement strength.
This supports further our quantum-correlation-based
understanding to the result in Fig.\ 3(a), since decoherence
indeed suppresses the quantum correlation, as shown in Fig.\ 4.

\section{Summary and Discussions}

We have reexamined the problem of quantum estimation of the
Rabi frequency
of qubit oscillations based on continuous measurement.
We specified our research to the superconducting circuit-QED system
which may provide an attractive platform for experimental examination.
Our central result is that, by proper design of the measurement strength,
the estimate precision can scale with the measurement time
beyond the standard quantum limit.
We understood this result by quantum correlation between the output signals
which is supported by checking the effect of quantum efficiency of the measurement.
Our conclusion is also supported by the scaling behavior
of the associated Fisher information, as shown in Appendix A.
We expect this preliminary result to inspire further studies
on this interesting problem,
including searching for better schemes of continuous measurement
and special techniques such as feedback and quantum smoothing.

As a final remark,
we mention again that the present work is an extension of the previous studies
on the quantum estimation of parameters by continuous measurements
\cite{Jac11,Molm13,Molm14,Molm15,Molm16,Jor17,Par17,Gen17}.
In particular, the effective measurement operator ($\sigma_z$) for the Rabi oscillation
is essentially the same as considered in Ref.\ \cite{Jor17},
where the main interest was focused on
accelerating the likelihood-estimation method and the estimation of drifting parameters.
This focus may have caused an overlook of the $T$ (measurement time)
scaling behavior of the estimate precision.
Probably affected by the $1/\sqrt{T}$ scaling concluded
by M\o lmer {\it et al} \cite{Molm13,Molm14,Molm15,Molm16},
this scaling was also briefly mentioned in Ref.\ \cite{Jor17}
below Eq.\ (16) associated with Fig.\ 2(c)
(despite that the result in Fig.\ 2(c) is a bit worse than the $1/\sqrt{T}$ scaling).
This difference, compared to our Fig.\ 3(a),
may originate from not finding a proper measurement strength
{\it and} simulating fewer number of trajectories there.
As a further support, in Appendix A we include
the result of our simulated Fisher information
and find similar scaling behavior beyond the SQL,
being consistent with that shown in Fig.\ 3(a).

Another point is that the measurement time we simulated
may be not long enough to reach the asymptotic behavior.
However, the scaling behavior even for this `intermediate' regime is
relevant in practical sense to the estimation problem under study.
We notice that the scaling behavior even for relatively short time
has been considered with interest.
For instance, in Refs.\ \cite{Gar13,Gar16}, it was found that
the quantum Fisher information can become quadratic in times
shorter than the correlation time of the dynamics.

\vspace{0.3cm}
{\flushleft\it Acknowledgements.}---
This work was supported by the
National Key Research and Development Program of China
(No.\ 2017YFA0303304) and the NNSF of China (No.\ 11675016).


\appendix

\section{Scaling Behavior of the Fisher Information}

In this Appendix, we carry out the Fisher information
associated with the present continuous measurement scheme.
The Fisher information is given by
\bea
{\cal F}_T(\Omega) = \int dx\, {\cal P}(x|\Omega)\left(
\frac{\partial\ln{\cal P}(x|\Omega)}{\partial\Omega} \right)^2  \,.
\eea
Here the short hand notation
$x=\{{\cal I}_1, {\cal I}_1,\cdots, {\cal I}_N\}$
is introduced for simplicity, and the integration
is in principle over all the possible output currents
from measurement realizations over time $T$.

In practice, we compute the Fisher information by
numerically averaging 20000 trajectories (realizations).
For each trajectory, we compute the derivative
$\partial\ln{\cal P}/\partial\Omega$
from the likelihood function at the real value $\Omega$.
In Fig.\ 5 we show the result of Fisher information against the measurement time $T$,
for the measurement strength $\Gamma_m/\Omega_R=0.25$ corresponding to Fig.\ 3(a).
In particular, in Fig.\ 5(b) we compare the result with the scaling behaviors
of the RMS variance $\delta\Omega$ and the SQL.
As in the plots of Figs.\ 3 and 4 in the main text,
here we plot $\sim 1/\sqrt{{\cal F}_T}$ by equating it
with the simulated RMS variance $\delta\Omega$ at the starting point.
Then, from this type of plotting and if we assume ${\cal F}_T\sim T^n$,
we can deduce the scaling index $n>1$, which exceeds the SQL scaling.
Moreover, in Fig.\ 5(b), we find satisfactory agreement between the $T$ scalings
of the Fisher information and the RMS variance $\delta\Omega$.
This supports further the conclusion we achieved in the main text.

\begin{figure}[!htbp]
  \centering
  \includegraphics[width=6.0cm]{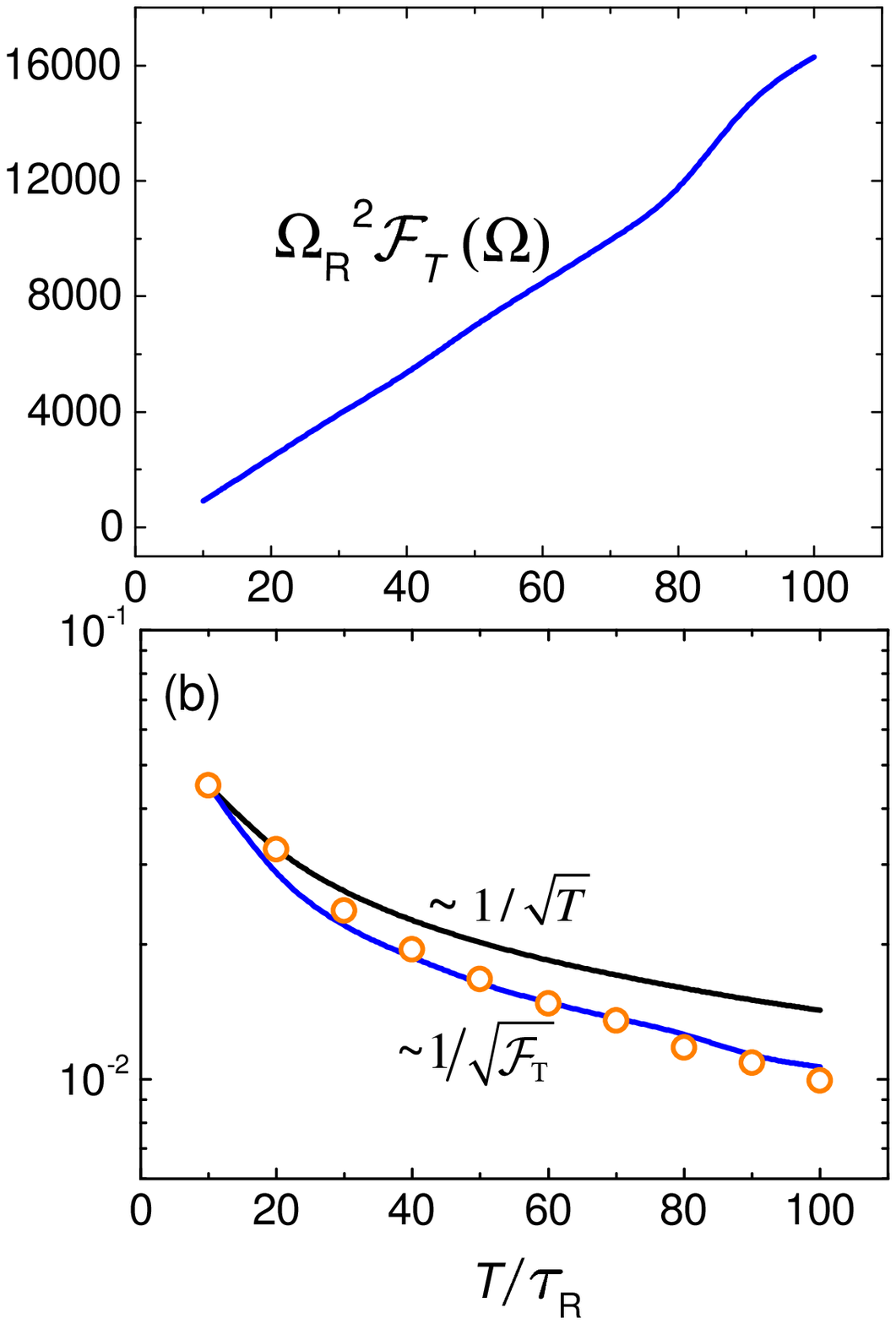}
  \caption{
Scaling behavior of the Fisher information ${\cal F}_T(\Omega)$
against the measurement time $T$,
for measurement strength $\Gamma_m/\Omega_R=0.25$ corresponding to Fig.\ 3(a).
Particularly, in (b) we compare the time scaling of the Fisher information
(blue curve) with the SQL (black curve)
and the RMS variance $\delta\Omega$ (orange circles, taken from Fig.\ 3(a)).  }
  \label{Fisher}
\end{figure}

\end{CJK*}
\end{document}